\documentclass[aps,superscriptaddress,showpacs,nofootinbib,eqsecnum]{revtex4}

\usepackage{epsfig}

\begin{document}

\title{Critical Dopant Concentration in Polyacetylene and Phase Diagram from a
  Continuous Four-Fermi Model}

\author{Heron Caldas} \email{hcaldas@ufsj.edu.br} \affiliation{Departamento de
  Ci\^{e}ncias Naturais. Universidade Federal de S\~{a}o Jo\~{a}o del Rei,
  36300-000, S\~{a}o Jo\~{a}o del Rei, MG, Brazil}

\author{Jean-Lo\"{\i}c Kneur} \email{kneur@lpta.univ-montp2.fr}
\affiliation{Laboratoire de Physique Th\'{e}orique et Astroparticules - CNRS -
  UMR5207 Universit\'{e} Montpellier II, France}

\author{Marcus Benghi Pinto} \email{marcus@fsc.ufsc.br}
\affiliation{Departamento de F\'{\i}sica, Universidade Federal de Santa
  Catarina, 88040-900 Florian\'{o}polis, SC, Brazil}

\author{Rudnei O. Ramos} \email{rudnei@uerj.br} \affiliation{Departamento de
  F\'{\i}sica Te\'orica, Universidade do Estado do Rio de Janeiro, 20550-013
  Rio de Janeiro, RJ, Brazil}

\begin{abstract}
  The Optimized Perturbation Theory (OPT) method, at finite temperature and
  finite chemical potential, is applied to the field theory model for
  polyacetylene.  The critical dopant concentration in {\it
    trans}-polyacetylene is evaluated and compared with the available
  experimental data and with previous calculations.  The results obtained
  within the OPT go beyond the standard mean field (or large-$N$)
  approximation (MFA) by explicitly including finite $N$ effects.  A critical
  analysis of the possible theoretical prescriptions to implement and
  interpret these corrections to the mean field results, given the available
  data, is given.  {}For typical temperatures probed in the laboratory, our
  results show that the critical dopant concentration is only weakly affected
  by thermal effects.

\end{abstract}

\pacs{64.60.A-,71.30.+h,11.10.Kk}

\maketitle

\section{Introduction}

About three decades ago a remarkable discovery was made that {\it
  trans}-polyacetylene (CH)$_x$ doped with halogens could behave as a metal,
exhibiting electrical conductivity~\cite{Shirakawa}. Since then several
striking features have been shown by conjugated polymers, such as electronic,
optical and magnetic properties, which give these materials a wide range of
applicability~\cite{Nature1,Nature2,Nature3,Lin1}. Before reaching the
metallic state, polyacetylene can be converted into a semiconductor, depending
on the concentration of dopant, $y$, defined as the number of doped electrons
per carbon atom. {}For lower dopant concentration, the conduction can be
described in terms of topological excitations, such as (spinless charged)
polarons, bipolarons or solitons~\cite{Review}.  Experiments have also shown
that the observed non-metal to metal transition in polyacetylene, that
typically happens when the dopant concentration is increased to a critical
value, $y_c \approx 6 \%$, corresponds to a first-order
transition~\cite{Fernando}.

{}From the theoretical point of view, the role of the (self-localized)
solitons in the charge-transport process is successfully described by the
model Hamiltonian proposed by Su, Shrieffer, and Heeger (SSH)~\cite{SSH}.  SSH
were primarily interested in the low-energy excitations of the one-dimensional
polymer. In this case, the electronic correlation length, $\xi$, becomes much
larger than the lattice constant, $a$, and an effective model Hamiltonian
incorporating only the electron-phonon coupling as an interaction term, and
neglecting electron-electron Coulomb repulsion, was shown to be able to give a
very good description of the system. To take into account electron-electron
interactions, extended versions of the SSH Hamiltonian, such as Hubbard
models~\cite{Lin1,Lin2}, have been later employed.

The continuum limit of the SSH Hamiltonian is known as the
Takayama--Lin-Liu--Maki (TLM) model \cite{tlm}, which is a field theory with
two-flavor Dirac fermions. The TLM model gives a very good approximation to
the discrete SSH model provided two conditions hold: the band gap (denoted as
$E_{\rm gap}$) between $\sigma$ electron bonding and anti-bonding states is
much smaller than the $\pi$-band width (denoted as $W$) and the typical
distance between the neighboring single-electron levels near the {}Fermi
energy is much smaller than the gap. The first condition is automatically
fulfilled for {\it trans}-polyacetylene, where from the typical measured
values $E_{\rm gap}/W \simeq (1.8 \, {\rm eV})/(10 \, {\rm eV}) = 0.18$
\cite{Review}, while the second one is met provided the application is for a
chain with a large size~\cite{Maxim}. At the same time, in many conducting
polymers the characteristic correlation length, $\xi$, is much larger than the
lattice constant, $a$. These two quantities are related to $E_{\rm gap}$ and
$W$ by $\xi/a = W/E_{\rm gap}$ \cite{Review}.

Some authors \cite{CB} have recognized that the corresponding Lagrangian of
the TLM model is analogous to a field theoretical model for fermions with
quartic self-interactions in 1+1 dimensions, also known as the Gross-Neveu
(GN) model \cite{gn}.  It is worth recalling that the large-$N$, or mean-field
approximation (MFA), predicts that, at $T=0$, the GN model in 1+1 dimensions suffers
a first-order phase transition at the critical chemical potential value,
$\mu_c = \Delta_0 /\sqrt{2}$, where $\Delta_0$ is the order parameter value at
$T=0=\mu$ \cite{wolff}.  Other works making use of the GN model applied to the
study of the properties of polyacetylene include Refs.
\cite{fradkin,otherrefs}.  However, to our knowledge, the issue of
investigating the phase diagram and temperature effects in the dopant
concentration in polyacetylene have not been addressed, in the context
of the GN model, so far.

Since we can relate the dopant concentration in terms of the chemical
potential, the GN model works as an analog model that can describe, in a field
theory language, the non-metal to metal phase transition observed in the {\it
  trans}-(CH)$_x$.  An early study of the transition from the (chiral symmetry
broken) low-density to the (symmetry restored) high-density phase, has been
performed by the authors of Ref. \cite{CM} using the GN model, by assuming
fermions with $N=2$ flavors, but considering only the leading order in a $1/N$
expansion. Despite this approximation, it was found an excellent agreement, of
about $6\%$, between the theoretical and experimental critical doping
concentration, $y_c$.  Later, the same authors have used a mixture of the
thermodynamic Bethe ansatz (TBA) and the $1/N$ expansion at the next to the
leading order (NLO) to evaluate finite $N$ corrections to $\mu_c$ \cite{CM2}.
Surprisingly, their NLO does not agree so well with the experimental results,
deviating by about $20 \%$. One of our main motivations in the present work is
to shed some light into this rather peculiar behavior.  After all, one expects
that finite $N$ corrections should improve convergence, especially in a case
where $N=2$, like it is the case in the theoretical model description of
polyacetylene.  With this aim we start by remarking that the critical dopant
concentration, $y_c$, is related to the critical density, $\rho_c$, and the
equilibrium space, $a$, between the $x$ coordinates of successive $CH$ groups
in the undimerized structure by $y_c= \rho_c a$.  However, in Ref. \cite
{CM2}, the authors did not compute the finite $N$ corrections to the critical
density $\rho_c$ but rather to the value of the chemical potential $\mu_c$.
Nevertheless, they concluded that, for $N=2$, $\rho_c(N)$ should change by
about $20 \%$ since it is proportional to $\mu_c(N)$, which changes by the
same amount. We believe that this conclusion is not necessarily right since
additional $N$ and/or coupling ($\lambda$) dependent factors, ${\cal
  H}(\lambda,N)$, could appear when obtaining $\rho$ by deriving the pressure
with respect to $\mu$ so that $\rho_c(N)$ could be proportional to ${\cal
  H}(\lambda, N)\, \mu_c(N)$ where, for example, ${\cal H}(\lambda, N \to
\infty) \to 1$ in the mean-field approximation. Second, $\rho$ depends on
parameter values such as $E_{\rm gap}$ and $W$ that sets $\xi/a =W/E_{\rm
  gap}$ but only the value $\xi/a \simeq 7$, which corresponds to the original
SSH parameters \cite {SSH} $E_{\rm gap} = 1.4\, {\rm eV}$ and $W = 10\, {\rm
  eV}$, was considered in Refs. \cite {CM,CM2}. We must recall that the
original SSH parameter values, especially the one regarding $E_{\rm gap}$,
have been further updated already in the Ref. \cite{Review}.

Here, we shall investigate the same problem considering again the 1+1
dimensional GN model, but not restricting the analysis to the large-$N$
approximation.  On general grounds, one may eventually wonder on the relevance
of going beyond the large $N$ (or equivalently mean field) approximation in
the present context, given that the equivalence between the original TLM model
and the GN field theory was established strictly speaking at the mean field
level: indeed, there are important phenomenological aspects and properties of
the polyacetylene that are anyway not correctly described by the sole TLM (for
a review on both the success and limitations of the TLM/GN equivalence see
e.g. \cite{campbell}). Nevertheless, it appears sensible here to push the
correspondence further in this direction at least because the TLM model is
effectively a two-flavor only model: since some results and techniques to go
beyond large $N$ are available, these can be expected to improve the
comparison of the GN model to polyacetylene data, as we shall see later.  In
this work we use the method of the optimized perturbation theory (OPT) \cite
{linear}, which is a nonperturbative method allowing to consider corrections
going beyond the large-$N$ (or mean-field) approximation. Our aim is to obtain
relations for $\rho_c$ which include finite $N$ corrections.  {}For this we
will reconsider Landau's free energy density for the 1+1 dimensional GN model
that has been recently derived by some of the present authors in Ref.
\cite{prdgn2}, where the OPT results for the phase diagram show how this
method correctly improves over mean-field results, in accordance with Landau's
theorem for phase transitions in one space dimensions. In the same reference
one finds $N$ dependent analytical equations for the order parameter and
chemical potential which turn out to be crucial for the present application. A
detailed description of how the method works can be found in Ref.
\cite{prdgn2} and references therein.
      
In the present application, we will indeed show that within the OPT one can
obtain the critical dopant concentration, $y_c$, in a consistent way,
producing results which are in very good agreement with the experimental
values for the {\it trans}-(CH)$_x$ when up-to-date parameter values are
considered.  {}For completeness however, we will examine several possible
theoretical prescriptions to implement and interpret these corrections to the
mean field results.  More precisely, as we shall see, the corrections to the
mean-field may be interpreted in different ways within the context of
polyacetylene. This is due to the fact that its description in terms of the
Gross-Neveu field theory model is strictly speaking only an approximation.  We
shall also pay attention to the non negligible experimental error on some of
the relevant available data.  Apart from considering finite $N$ corrections to
$y_c$, we also estimate how thermal effects affect this quantity.

It is worth mentioning that an extension to the $2+1$ dimensional GN theory is
straightforward and may be of use in the investigation of superconducting
electrons in quasi-two-dimensional systems~\cite{Marino}. In connection with
the 2+1 dimensional GN model, the OPT has allowed \cite{prdgn3} to redraw its
phase diagram leading to the precise location of a tricritical point in the
$T-\mu$ plane and the discovery of a mixed ``liquid-gas" phase that remained
undetermined for almost twenty years, since the first mean-field results
appeared \cite{klimenko}. The compatibility of the OPT and Landau's expansion
for the free energy density has been shown in Ref. \cite{plb}.  In the
condensed matter domain, higher order OPT results have considerably improved
early applications of the same method \cite{prb} producing some of the most
precise analytical values for the shift of the critical temperature, $\Delta
T_c$, of an interacting homogeneous Bose gas when compared to the ideal gas
\cite {pra}. {}Finally, the OPT convergence has also been proved in connection
with critical theories \cite{prl,braaten}.

As far as thermal effects in the GN model at one space dimension and the
application of these results to polyacetylene are concerned, a few comments
are appropriate here. We recall that due to a well known no-go theorem
\cite{landau,mermin,coleman1}, for a one-dimensional system at any finite
temperature we should expect no phase transition related to a discrete
symmetry breaking (in this case a discrete chiral symmetry in the massless GN
model considered in this work).  This is due to kink-like inhomogeneous
configurations \cite{ma} that come to dominate the action functional, instead
of just homogeneous, constant field configurations. This is to be contrasted
to the phase transition observed since long ago in the GN model in one space
dimension in the mean-field, large-$N$ approximation \cite{wolff}. This result
is explained by the way the thermodynamic and the mean-field approximation are
performed.  If the thermodynamic limit is taken before the mean-field
approximation, those large nonhomogeneous fluctuations dominate and the
theorem is observed. However, if the mean-field approximation is considered
first, the fluctuations are suppressed, thus seem to evade the no-go theorem.
Since we are here applying the GN model as an effective analog model for the
polyacetylene and this is in practice a finite size system, we do not expect
the theorem to be completely observed here.  In fact, a phase transition at
finite temperatures is indeed observed and measured in the laboratory.
Nevertheless, polyacetylene is a well-known system exhibiting a rich spectrum
of nontrivial fluctuations, from solitons to polaron excitations
\cite{Review}.  Therefore, we may expect not only homogeneous like
configurations (like in the mean-field approach), but also that the inclusion
of these excitations in any theoretical calculation in this model should be
considered.  In this context, for example in the GN field theory model, by
accounting for kink-like configurations in the large-$N$ approximation, the
authors of Ref.  \cite{gnpolymers} found evidence for a crystal phase that
shows up in the extreme $T \sim 0$ and large $\mu$ part of the phase diagram,
while the other extreme of the phase diagram, for large $T$ and $\mu \sim 0$,
seemed to remain identical to the usual large-$N$ results for the critical
temperature and tricritical points, which are well known results \cite{wolff}
for the GN model.  In this work we will only consider homogenous vacuum
backgrounds in our thermodynamical calculations applied to the polyacetylene.
By comparing our results with the experimental ones we can roughly estimate
the importance of soliton-like excitations in the system.  {}From our results,
we estimate that these effects are expected to be small in the context of
applying the GN model as an effective analog model to describe the
thermodynamics of polyacetylene at low (laboratory) temperatures.

This work is organized as follows. In Sec. II we briefly present the TLM model
and its relation to a four-fermion theory, which can be identified as the GN
model.  In Sec. III we review the computation of the free energy for the GN
model by using the OPT method. In the same section, the temperature dependent
density is obtained. The gap equation is used to set up the parameter values
in Sec. IV.  In Sec. V we show our phase diagrams for the {\it
  trans}-polyacetylene (CH)$_x$ both in the $T-\mu$ and $T-\rho$ planes.  The
critical dopant density, at zero and finite temperatures, is considered in
Sec. VI. Our concluding remarks are given in Sec. VII.

\section{The Takayama--Lin-Liu--Maki and Gross-Neveu Models}

The Takayama--Lin-Liu--Maki (TLM) Hamiltonian is the continuum version for the
original SSH model and it is given in terms of a fermionic field, $\psi$, and
a scalar field, $\Delta$, representing the coupling of the electron gas to the
local value of the dimerization and it is expressed by the Hamiltonian
\cite{tlm}

\begin{equation}
H_{\rm TLM} = \frac{1}{2 \pi \hbar v_F \lambda_{\rm TLM}}
\int dx \Delta^2(x) + \sum_s \int d x \psi^\dagger (x)
\left[ - i \hbar v_F \sigma_3 \partial_x + \sigma_1 \Delta (x)
\right] \psi(x)\;,
\label{HTLM}
\end{equation}
where the sum is over the spin states, $\sigma_i$ are the Pauli matrices, $v_F
= k_F/m$ is the Fermi velocity and $\lambda_{\rm TLM}$ is a dimensionless
coupling defined by

\begin{equation}
\lambda_{\rm TLM} = \frac{ 2 \alpha^2}{\pi t_0 K}\;,
\label{lambdaTLM}
\end{equation}
where $\alpha$ is the $\pi$-electron-phonon coupling constant of the original
SSH Hamiltonian, $K$ is the elastic chain deformation constant and $t_0$ is
the hopping parameter, which is expressed in terms of the Fermi velocity and
the equilibrium space $a$ between the $x$ coordinates of successive $CH$
groups in the undimerized structure as $t_0 = \hbar v_F/(2\,a)$.

Note from Eq. (\ref{HTLM}) that a nonvanishing (constant) value for $\Delta$
leads to a mass term for the fermions, thus breaking the chiral symmetry
exhibited by $H_{\rm TLM}$ and opening an electronic energy gap in the system.
The presence of a gap prevents electrons to move to the conduction band and,
thus, the system effectively behaves as a non-metal.  The effect of the
addiction of dopants to the system is to decrease the electronic energy gap,
till it vanishes at some critical dopant concentration and the system starts
to behave as a metal. In general, a kinetic term for the scalar field emerges
when taking the continuum limit of the SSH model. However, we consider the
usual adiabatic approximation of neglecting the lattice vibrations, valid for
energies for the optical-phonons (given by $\hbar \omega_0$) smaller than the
gap magnitude ($2 \Delta$).  In particular, for typical values found for
polyacetylene \cite{Review}, $2 \Delta \approx 1.8 \, {\rm eV}$ and $\hbar
\omega_0 \approx 0.12 \; {\rm eV}$, this is regarded as a valid approximation.

The model described by Eq. (\ref{HTLM}) can easily be shown to correspond to a
four-Fermi model if we eliminate the scalar field $\Delta$ from Eq.
(\ref{HTLM}), e.g. by using its equation of motion. Then, putting the TLM
model in the Lagrangian density form one obtains

\begin{equation}
{\cal L}_{\rm TLM} =
-\frac{1}{2 \pi \hbar v_F \lambda_{\rm TLM}} 
\Delta^2 + \psi^\dagger
\left( i \hbar \partial_t - i \hbar v_F \gamma_5 \partial_x - \gamma_0
\Delta 
\right) \psi\;,
\label{LagTLM}
\end{equation}
where we have identified $\gamma_5 = - \sigma_3$ and $\gamma_0 = \sigma_1$.
Now, eliminating $\Delta$ from Eq. (\ref{LagTLM}) upon using $\gamma_1 = i
\sigma_2$, as well as the usual relations between the Dirac matrices, leads to

\begin{equation}
{\cal L}_{\rm TLM} =
\bar{\psi}
\left( i \hbar \gamma_0 \partial_t - i \hbar v_F \gamma_1 \partial_x
\right ) \psi +
\frac{\lambda_{\rm GN}}{2N}\hbar v_F \left(\bar{\psi}\psi\right)^2\;,
\label{LagTLM-GN}
\end{equation}
which is just a four-Fermi Lagrangian density corresponding to the massless GN
model \cite{gn} where $N$ is the number of fermion flavors ($N=2$ for
polyacetylene).  In Eq. (\ref{LagTLM-GN}) we have used Eq.  (\ref{lambdaTLM})
to define the GN coupling as

\begin{eqnarray}
\lambda_{\rm GN} = N \pi \lambda_{\rm TLM}
= \frac{2 N   \alpha^2}{t_0 K}\;.
\label{lambdaGN}
\end{eqnarray}

\section{The Gross-Neveu Model in the Optimized Perturbation Theory}

Let us now turn our attention to the implementation of the OPT procedure
\cite{linear} (for a long, but far from complete list of references, please
see also Ref. \cite{prdgn2} and references in there) within the model
Lagrangian density given by Eq. (\ref{LagTLM-GN}).  Applying the usual OPT
interpolation prescription to the {\it original} four-Fermi theory, Eq.
(\ref{LagTLM-GN}), we define the interpolated theory

\begin{equation}
{\cal L}_{\delta}(\psi, {\bar \psi}) =
\bar{\psi}
\left( i \hbar \gamma_0 \partial_t - i \hbar v_F \gamma_1 \partial_x
\right ) \psi  -
\eta (1-\delta) {\bar \psi} \psi
+ \delta \frac {\lambda_{\rm GN}}{2N} \hbar v_F({\bar \psi} \psi)^2\;,
\label{GNlde}
\end{equation}

\noindent
where $\eta$ is an arbitrary mass parameter.  It is easy to verify that at
$\delta=0$ we have a theory of free fermions, and the original theory is
recovered for $\delta=1$. Now, by re-introducing the scalar field $\Delta$,
which can be achieved by adding the quadratic term (corresponding to a
Hubbard-Stratonovich transformation)

\begin{equation}
- \frac{ \delta N}{2 \hbar v_F \lambda_{\rm GN}} \left ( \Delta +
\frac {\lambda_{\rm GN}}{N} \hbar v_F {\bar \psi} \psi \right )^2 \,,
\end{equation}
to ${\cal L}_{\delta}(\psi, {\bar \psi})$, one obtains the interpolated model
corresponding to the original TLM model given by Eq.  (\ref{LagTLM}),

\begin{equation}
{\cal L}_{\delta} =
\bar{\psi}
\left( i \hbar \gamma_0 \partial_t - i \hbar v_F \gamma_1 \partial_x
\right ) \psi -
\delta \Delta {\bar \psi} \psi - \eta (1-\delta) {\bar \psi} \psi
- \frac {\delta N }{2 \hbar v_F \lambda_{\rm GN} } \Delta^2   \;.
\label{GNdelta}
\end{equation}

\noindent 
Since Eq. (\ref{GNdelta}) is the same model already studied in Ref.
\cite{prdgn2}, so we do not repeat all the details related to the free energy
density derivation here, where only the main steps and results relevant for
our application to the polyacetylene will be presented.

Generally, the OPT method can be implemented as follows. Any physical
quantity, $\Phi^{(k)}$, is {\it perturbatively} computed from the interpolated
model, up to some finite order-$k$ in $\delta$, which is formally used only as
a bookkeeping parameter and set to the unity at the end of calculation.  But
in this process any (perturbative) result at order $k$ in the OPT remains
$\eta$ dependent.  This arbitrary (a priori) parameter is then fixed by a
variational method that then generates nonperturbative results, in the sense
that it resums to all orders a certain class of perturbative contributions
through self-consistent equations. Such optimization method is known as the
principle of minimal sensitivity (PMS) and amounts to require that
$\Phi^{(k)}$ be evaluated at the point where it is less sensitive to this
parameter.  This criterion translates into the variational relation \cite{pms}

\begin{equation} 
\frac {d \Phi^{(k)}}{d \eta}\Big |_{ \delta=1, \eta=\bar \eta} = 0 \;.
\label{PMS} 
\end{equation}

\noindent
The optimum value $\bar \eta$ that satisfies Eq. (\ref{PMS}) must be a
function of the original parameters, including the couplings, thus generating
``non-perturbative" results.  In our case, we are interested in evaluating the
optimized free energy at finite temperature and density for the scalar field,
$\Delta$, once the fermions have been integrated out.

\subsection{The Optimized Free Energy Density}

To order-$\delta$, Landau's free energy density (or effective potential, in
the language of quantum field theories) was evaluated in Ref. \cite{prdgn2}
using functional and diagrammatic techniques. The result is

\begin{eqnarray}
{\cal F} (\Delta_c, \eta, T, \mu) &=&
\delta \frac {N \Delta_c^2}{2 \lambda_{\rm GN} v_F \hbar} -
 \frac{N}{2\pi v_F \hbar } \left \{ \eta^2 \left [ \frac {1}{2} + \ln \left (
\frac {M}{\eta} \right ) \right ] + 2 (kT)^2 I_1(\eta,\mu,T) \right \}
\nonumber \\
&+& \delta \frac{N\eta(\eta-\Delta_c)}{\pi v_F \hbar}
\left[\ln\left(\frac{M}{\eta}\right) - I_2 (\eta,\mu,T) \right]
\nonumber \\
&+& \delta \frac { \lambda_{\rm GN}}{4\pi^2 v_F \hbar} \left \{ \eta^2
  \left [ \ln \left ( \frac {M}{\eta}  \right ) - I_2(\eta,\mu,T)
\right ]^2 + (kT)^2 I^2_3(\eta,\mu,T) \right \}\;.
\label{Vdelta1}
\end{eqnarray}
where $k$ is the Boltzmann constant and the functions $I_1$, $I_2$ and $I_3$
are given respectively by

\begin{equation}
I_1(\eta,\mu,T) = \int_0^\infty dx
\left\{ \ln \left[ 1+
e^{-\sqrt{x^2+\eta^2/(kT)^2}-\mu/(kT) } \right]
+ \ln \left[ 1+e^{-\sqrt{x^2+\eta^2/(kT)^2}+\mu/(kT) } 
\right] \right\}\;,
\label{I1}
\end{equation}

\begin{equation}
I_2(\eta,\mu,T)=\int_0^\infty \frac {d x}{\sqrt{x^2 +\eta^2/(kT)^2}} \left[
\frac{1}{e^{\sqrt{x^2+\eta^2/(kT)^2}+\mu/(kT) }+1}+
\frac{1}{e^{ \sqrt{x^2+\eta^2/(kT)^2}-\mu/(kT)  }+1} 
\right] \;,
\label{I2}
\end{equation}
and

\begin{eqnarray}
I_3 (\eta,\mu,T)
= \sinh\left(\frac{\mu}{kT}\right) \int_0^\infty d x
\frac{1}{\cosh\left(\sqrt{x^2 + \eta^2/(kT)^2 }\right) + 
\cosh\left[\mu/(kT)\right]}  \;.
\label{I3}
\end{eqnarray}
In Eq. (\ref {Vdelta1}), $\Delta_c$ is a constant field configuration for the
scalar field and $M$ is an arbitrary energy scale introduced during the
regularization process used to compute the appropriate momentum integrals.  In
the computation performed in Ref. \cite {prdgn2}, the free energy density has
been renormalized using the $\overline {\rm MS}$ scheme for dimensional
regularization.  We also note that Eq.  (\ref{Vdelta1}), evaluated at first
order in the OPT, already takes into account corrections beyond the large-$N$
result.

By optimizing Eq. (\ref{Vdelta1}) through the PMS condition, Eq. (\ref{PMS}),
we obtain the optimum value, $\bar{\eta}$, for the mass parameter, which is
then re-inserted back in Eq. (\ref{Vdelta1}), allowing us to compute the order
parameter ${\overline \Delta}_c$ that minimizes the free energy. {}Using the
PMS procedure we then obtain, from Eq.  (\ref{Vdelta1}), the general
factorized result \cite{prdgn2}

\begin{eqnarray}
\left \{ \left [ {\cal Y}(\eta,\mu,T)
+ \eta \frac {d}{d \eta} {\cal Y}(\eta,\mu,T) \right ]
\left[ \eta - \Delta_c +
\eta \frac{\lambda_{\rm GN}}{2 \pi N} {\cal Y}(\eta,\mu,T) \right]
+ \frac{(kT)^2 \lambda_{\rm GN}}{2 \pi N}I_3(\eta,\mu,T)
\frac {d}{d \eta}I_3(\eta,\mu,T)
\right\}\Bigr|_{\eta = \bar{\eta}} = 0 \;,
\label{genpms}
\end{eqnarray}
where we have defined the function

\begin{equation}
{\cal Y}(\eta,\mu,T) =\ln \left ( \frac{M}{\eta} \right ) -
I_2(\eta,\mu,T) \;.
\end{equation}
Considering the $\lambda_{\rm GN}/N$ dependent solution one notices that, when
$N \to \infty$ in Eq. (\ref{genpms}), ${\bar \eta} = \Delta_c$ and the
mean-field standard result is exactly reproduced as usual \cite{prdgn2,npb}.
{}For finite $N$, as is our interest here, ${\bar \eta}$ and $ \Delta_c$ have
to be found self-consistently by solving the gap equation $d{\cal F}/d
\Delta_c=0$ and the PMS equation $ d{\cal F}/d \eta=0$ \cite{prdgn2}.

\subsection{ The density at finite temperature}

The thermodynamical potential (per volume) is defined as the free energy
density at its minimum, $\Omega(T,\mu) = {\cal F}({\bar \eta}, {{\overline
    {\Delta}}_c,T, \mu})$ and the pressure follows as $P(T,\mu) = -
\Omega(T,\mu)$. The density is then obtained by the usual relation $\rho = d
P/d\mu$. We must also recall that $d{\cal F}/d \Delta_c=0$ at
$\Delta_c={\overline \Delta}_c$, due to the gap equation, and that $ d{\cal
  F}/d \eta=0$ at $\eta={\bar \eta}$, due to the PMS equation. Then, terms
like $(d{\cal F}/d \Delta_c)(d\Delta_c/d \mu)$ and $(d{\cal F}/d \eta)(d\eta/d
\mu)$ do not contribute.  One then obtains

\begin{eqnarray}
\rho (T,\mu)&=& \frac {1}{v_F \hbar} \left [ (kT)^2 \frac{N}{\pi }
I_1^\prime(\eta,\mu,T)
+ \eta (\eta -\Delta_c) \frac{N}{\pi} I_2^\prime(\eta,\mu,T)
\right. \nonumber \\
&-& \left.  \frac{\lambda_{\rm GN}}{2 \pi^2} \eta^2 I_2(\eta,\mu,T)
I_2^\prime(\eta,\mu,T)-
\frac{\lambda_{\rm GN}}{2 \pi^2} (kT)^2 I_3(\eta,\mu,T)
I_3^\prime(\eta,\mu,T) \right ]
\Bigr|_{\eta = \bar{\eta}, \Delta_c = {\overline \Delta}_c}
\label{density}
\end{eqnarray}
where the primes indicate derivatives with respect to $\mu$. This result will
be considered later when we investigate thermal effects in $y_c$.

\section{ The Gap Energy and Parameter set at $T=0$ and $\mu=0$}

In order to perform a numerical analysis we must fix all parameters. This can
be done by considering the gap energy.  In the GN language the order parameter
${\overline \Delta}_c$ is just the TLM gap parameter which, at $T=0$ and
$\mu=0$, we denote as $\Delta_0$.  At order-$\delta$ this quantity is given by
\cite{prdgn2}

\begin{equation}
 \Delta_0 =
M \exp
\left \{ -\frac {\pi}{\lambda_{\rm GN}\left ( 1- \frac{1}{2  N} \right ) }
\right \} \left( 1- \frac{1}{2  N} \right)^{-1}\;,
\label{deltaopt}
\end{equation}
where $M$ is an arbitrary (at the moment) renormalization scale to be
discussed further below.  Eq. (\ref{deltaopt}) explicitly includes corrections
beyond large-$N$, as obtained from our OPT approach.  More precisely, taking
the mean-field approximation, $N\to \infty$ in Eq.  (\ref{deltaopt}) and using
the relation $\lambda_{\rm GN}=N \pi \lambda_{\rm TLM}$, the OPT result
exactly recovers the mean-field result for $N=2$ \cite {Review},

\begin{equation}
 \Delta_{MF} =
M \exp [-1/(2 \lambda_{\rm TLM})]\;,
\label{deltaopt2}
\end{equation}
as one expects \cite{prdgn2,npb}.

Now some remarks concerning the arbitrary energy scale, $M$, and more
generally on the interpretation of Eq.(\ref{deltaopt}) in the present
polyacetylene context are useful.  Usually, in a renormalizable quantum field
theory, one can choose arbitrary value for $M$ and $\lambda_{\rm GN}$ will run
with the scale appropriately, at a given perturbative order, so that
$\Delta_0$ remains scale-invariant as dictated by the renormalization group.
{}For the above gap equation this means that

\begin{equation}
\frac {1}{\lambda_{\rm GN}(M)} = \frac {1}{\lambda_{\rm GN}(M_0)} +
\frac{\left( 1- \frac{1}{2  N} \right)}{\pi} 
\ln \left (
\frac {M}{M_0} \right )\,,
\label{RG}
\end{equation}
where $M_0$ is some reference (input) scale \footnote{Note that the
  $N$-dependence of the running coupling as dictated by Eq. (\ref{RG}) differs
  from the standard RG one, which (at one-loop order) has a coefficient given
  by $1-1/N$ instead of $1- 1/(2 N)$. This difference is a result of OPT which
  modifies standard one-loop order results, in a way expected to improve the
  latter.}.  Equations (\ref{deltaopt}), or Eq. (\ref{RG}), indicates that
$\lambda(M)$, or equivalently $\Delta_0$, is the only parameter to be fixed.
These equations also show that $\lambda(M) \to 0$ as $\ln^{-1}(M/M_0)$ when $M
\to \infty$ which is nothing else than the asymptotic freedom displayed by the
GN model.  However, in the polymer physics case, the interpretation is
somewhat different mainly because {\em both} $\Delta_0$ and the coupling
$\lambda_{\rm GN}$ are measurable quantities, as we shall exploit below.
Moreover in contrast to the renormalizable field theory case all quantities
here are expected to be directly finite, {\it i.e.}, without need of
renormalization due to the explicit high energy cutoff $\Lambda$ provided by
the $\pi$-band width, {\it i.e.}, $\Lambda \sim W$. While in our calculation
we have used dimensional regularization and renormalization mainly for
convenience, Eq.(\ref{deltaopt}) should be interpreted as giving the {\em
  finite}, $N$-dependent, corrections to the large-$N$ results, with the
arbitrary scale $M$ (originating from dimensional regularization) to be traded
for an explicit cutoff: $M\equiv \Lambda$ of order $\Lambda \sim W$. Now since
$M$ is a parameter from the theory, its precise value is thus a matter of
choice to some extent, as it does not need to coincide exactly with the
experimental parameter $W$.  This implies in particular that the scale $M$ can
be dealt with in alternative ways as we shall discuss next.

{}Consequently, we can consider different possible prescriptions for the basic
parameters of the problem, given also that some data appear to have non
negligible experimental uncertainties.

\begin{itemize}
  
\item{i)} {}First, in the prescription we label (I), $\lambda_{\rm GN}(M)$ can
  be simply set to its phenomenological value given by

\begin{eqnarray}
\lambda_{\rm GN}(M) =  \frac{8 N   \alpha^2}{W K}\;,
\label{lambdaGN2}
\end{eqnarray}
where we have used the relation $4 t_0 = W$. As discussed above this
implicitly defines a scale $M$ once assuming the theoretical prediction of Eq.
(\ref{deltaopt}).

{}Regarding the data numerical values, we note that this has been debated for
long and different set of values appeared in the literature (see e.g. Refs.
\cite{SSH,Review}).  As far as we are aware, it appears \cite{campbell}
however that the present widely accepted data values are: $K= 21$ eV/$\AA^2$,
$\alpha = 4.1 \, {\rm eV}/\AA$, $2\Delta_0 = 1.4-1.8 \, {\rm eV}$ and $W
\equiv 4t_0= 10 \, {\rm eV}$, which are essentially the conventional SSH
values \cite{SSH} except for possible higher values of $\Delta_0$
\cite{Review}, which appears as the less accurately determined experimental
input. Consequently, in our study we shall take this set of input but taking
the two extreme values of $\Delta_0$, that we will call set A and B,
respectively for $2 \Delta_0=1.4 \; (1.8) \;{\rm eV}$.

Comparing thus set A and B appears to us as a very conservative way of taking
into account those experimental uncertainties, although the higher value of
$2\Delta_0 \sim 1.8 \, {\rm eV}$ appears to be much more favored in the recent
literature.  Since all relevant physical quantities (such as the critical
density) will depend on the cutoff scale, $M$, as already mentioned one
possible prescription is to use Eq. (\ref{deltaopt}) to fix the cutoff $M$
value for given $\lambda_{\rm GN}(M)=8N \alpha^2/(WK)$ and $\Delta_0$ within
accuracy, {\it i.e.}, for each data set A and B.  Eq.  (\ref{deltaopt}) shows
that due to the presence of $N$ dependent constant term $1- 1/(2 N)$, the OPT
and the MFA will 
predict in this way different $M$
values even when using the same set of input data parameters. \\

\item{ii)} Another possible prescription, that we dub (II), is to set $M=W$
  exactly, cutting off the spectrum at an energy scale of $ - W/2$ \cite
  {Review,campbell}. In such case $\lambda_{\rm GN}(M=W)$ does not exactly
  match the experimental value as predicted by Eq.  (\ref {RG}). Most previous
  authors appear to haven chosen this prescription, {\it i.e.}, changing the
  coupling $\lambda_{\rm GN}$ value in order that the mean field model best
  fits the polyacetylene data.  We find however equally motivated to use the
  first interpretation since, as already mentioned, the polyacetylene data
  provides us with a rather precise $\lambda_{\rm GN}$ value, while there is
  some intrinsic arbitrariness in the precise cutoff scale value (equivalently
  in this case, $\Delta_0$ fixes the energy cutoff scale $M$ within some
  accuracy) \footnote{Note also that Eq. (\ref {RG}) also allows for any other
    intermediate prescriptions in which $ M \ne W $ and $\lambda_{\rm GN}(M)
    \ne 8N \alpha^2/(WK)$, but the simultaneous equalities for these
    quantities is excluded because of the $2\Delta_0=1.4-1.8$ eV values.}.
  Actually the two prescription are not fundamentally different: in the first
  one uses the arbitrariness of the cutoff to fit the data $\Delta_0$ and
  $\lambda_{\rm GN}$, while in the second one forces the coupling to fit the
  two scales $\Delta_0$ and $W$, but this is essentially translating the
  arbitrariness of the scale inside the
  exponential of Eq. (\ref{deltaopt}). \\
  
\item{iii)} {}Finally let us consider yet another possible prescription (or
  rather interpretation) of the GN model/polyacetylene data connexion.  It
  will define our prescription III.  Namely, bearing in mind that the
  equivalence between the original TLM and continuous GN model was strictly
  established only at the mean field theory level, we may redefine our OPT
  corrections in the framework of an {\em effective} mean field (EMF) GN
  description: more precisely, the OPT-modified gap energy Eq.
  (\ref{deltaopt}) can be fitted by the corresponding mean field expression
  Eq.  (\ref{deltaopt2}) provided that ones redefines ``effective" mean field
  coupling\footnote{Note that the meaning of effective coupling here is purely
    phenomenological, as obtained from a fit, and thus unrelated to the usual
    effective coupling $\lambda_{\rm GN}(M)$ as above discussed having the RG
    behavior.}  $\lambda^*_{EMF}$ and cutoff $W^*_{EMF}$:

\begin{equation}
 \lambda^*_{EMF} \equiv \lambda_{\rm GN} \left( 1- \frac{1}{2  N} 
\right)\;,\;\;\;\;
W^*_{EMF} \equiv M \left( 1- \frac{1}{2  N} \right)^{-1} \;, 
\label{defeff}
\end{equation}
together with the identification of these effective parameters to the measured
data.

\end{itemize}

{}This freedom of prescriptions, as discriminated above as prescriptions I, II
and III, actually reflects that neither the MFA nor the OPT-improved
expression of the gap energy are expected to be exact results. If available,
an exact, truly non-perturbative calculation of the gap energy would be
expected to fit nicely the three independent experimental measurements, $t_0$
(equivalently $W$), $\Delta_0$, and $\lambda_{\rm GN}$ (of course up to
limited experimental accuracy). Therefore for completeness and comparison
purpose, we will consider in the numerical results all these prescriptions
together with data sets A and B.  A summary of the different $M$ and
$\lambda_{\rm GN}$ values for each prescription and data set is given in Table
\ref{tabdata}.

Inspection of Table I indeed indicates rather different values of the ``bare"
coupling $\lambda_{\rm GN}$ for the three prescriptions, which is essentially
due to the large uncertainty in $\Delta_0$ between sets A and B. One should
not conclude from this that our description is lacking prediction.  In fact,
as we shall see later, the predictions for our main result on the critical
dopant estimate are not strongly dependent on the coupling values, and will be
only slightly different for the three cases (provided one uses the same
experimental data input).  Again, the most important variation will be due to
the large uncertainty on $\Delta_0$.

\begin{table}[htb]
{\begin{tabular}{c||c|c|c|c|c|c}
\hline 
 {}& I.A & I.B & II.A & II.B & III.A & III.B 
\\ \colrule \colrule
$\lambda^{\rm OPT}_{\rm GN}$ & 1.28 & 1.28  &  1.42 &  1.55 & 1.89 & 2.07\\ 
\colrule
 $\lambda^{\rm MFA}_{\rm GN}$& 1.28 & 1.28   &  1.18 & 1.30 & 1.18 & 1.30\\ 
\colrule
$M^{\rm OPT} (\rm eV)$ & 13.85 & 17.80 & 10 & 10  & 7.5 & 7.5 \\ \colrule
$M^{\rm MFA}(\rm eV)$ & 8.15 & 10.47 & 10 & 10 & 10  & 10 \\ 
\hline 
\end{tabular}}
\caption{\label{tabdata} The OPT and MFA values for $\lambda_{\rm GN}$ 
and $M$ obtained, from the gap equation, for parameter sets A 
($2 \Delta_0 = 1.4$ eV)  and B ($2 \Delta_0 = 1.8$ eV),
using prescriptions I--III for fixing the relevant parameters of the
model. The common values for both cases are: $K= 21$ eV/$\AA^2$, 
$\alpha = 4.1 \, {\rm  eV}/\AA$ and $W \equiv 4t_0= 10 \, {\rm
  eV}$. }
\end{table}

\section{Phase Diagrams}

\begin{figure}[htb]
  \vspace{0.5cm}
  \epsfig{figure=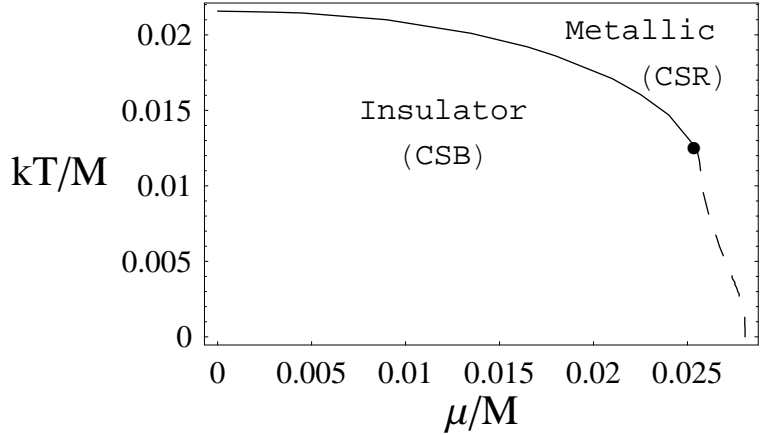,angle=0,width=10cm}
\caption[]{\label{Tmu} The OPT phase diagram, in the $kT/M$-$\mu/M$ plane,
  for $N=2$ and $\lambda=1.28$ (prescriptions IA and IB in Table
  \ref{tabdata}). The continuous line represents the second order transition
  whereas the dashed line represents the first order transition and the dot
  indicates the tricritical point, which occurs ate $kT/M \simeq 0.012 $ and
  $\mu/M \simeq 0.025$.}
\end{figure}

Having set the parameters for different prescriptions we can investigate the
phase diagrams for the theory. Let us start by locating the second order and
first order transition lines in the $T-\mu$ plane. This is shown in {}Fig.
\ref{Tmu} for the choice of prescription IB. It shows the appearance of a
tricritical point around $kT/M \simeq 0.012 $ and $\mu/M \simeq 0.025$.  Those
numbers would slightly change for the other prescriptions, with the overall
behavior qualitatively very similar.  Although the appearance of a tricritical
point is an interesting issue when considering the GN as a toy model for QCD,
it has no practical implications for the polyacetylene, in which case one is
concerned with temperatures lower than about $T_d \sim 400 \, K$, above which 
the polyacetylene is unstable and decomposes when heated (instead of melting) 
\cite{roth}, that is $k T_d/M \approx 0.0020$ for $M \approx 17.80 \,
{\rm eV}$ (prescription IB).  This type of phase diagram has been extensively
studied in Ref. \cite{prdgn2}, where analytical expressions for $T_c$,
$\mu_c$, as well as useful relations among tricritical points relations can be
found.

\begin{figure}[htb]
  \vspace{0.5cm}
  \epsfig{figure=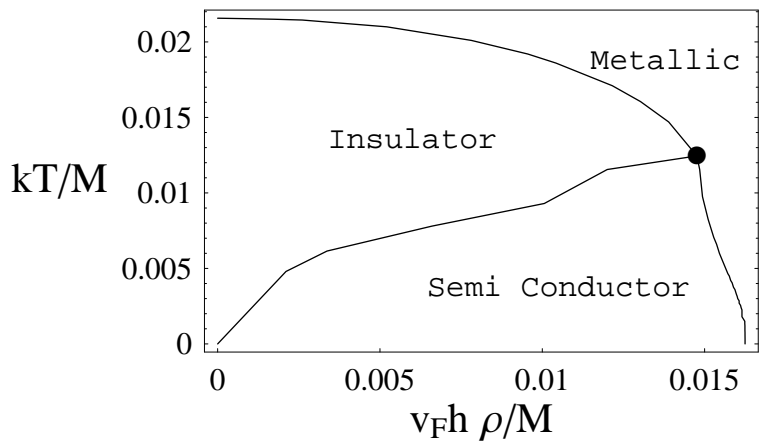,angle=0,width=10cm}
\caption[]{\label{rhoT} The  OPT phase diagram in the $kT/M-v_F \hbar\rho/M$ 
  plane for $N=2$ and $\lambda=1.28$ (prescriptions IA and IB in Table
  \ref{tabdata}).  The insulator region is associated with the unsymmetric
  (dimerized) phase, while the metallic region is associated with the
  symmetric (undimerized) phase.}
\end{figure}

As emphasized in the introduction, one of our goals here regards the
evaluation of the critical concentration, $y_c$, as a function of the
temperature.  With that aim one benefits from analyzing the phase diagram in
the $T-\rho$ plane since $y_c(T)$ is directly proportional to $\rho_c(T)$.
This is shown in {}Fig. \ref{rhoT}.  The dot in {}Fig. \ref{rhoT} indicates
the tricritical point above which the transition is of the second kind.  The
mixed (semi conductor) region is associated to the first order phase
transition. At $T=0$ the critical dopant density is approximately $\rho_c(0)
\simeq 0.016\, M/(v_F \hbar)$.  {}Figure \ref{rhoT} shows the situation where
the first order transition line, which appears in the $T-\mu$ plane, splits
into two lines limiting a coexistence, mixed (semiconducting) region. We note
that there are indeed experimental indications of a mixed phase for
polyacetylene for concentrations below the critical one \cite{mixed}.  It is
also interesting to note, from the same figure, that when one evaluates $y_c$
at $T=0$ using the GN model \cite{CM,CM2} the only observed transition is from
the semiconducting phase to the metallic one.  However, even at room
temperature (roughly $k T/M \approx 0.0015$) our figure displays another
transition from the (unsymmetric) insulator phase to the (mixed) semiconductor
phase which happens at a rather very small density, of the order $v_F \hbar
\rho/M \sim 10^{-5}$.  {}For this transition the critical density increases
with the temperature. On the other hand, the critical density when going from
the (mixed) semiconducting phase to the (symmetric) metallic one seems to
slightly decrease for low values of $T$.  {}Figure \ref{rhoT} also shows that
above $\rho_c(0)$, computed in the next section, the material is a conductor
at any temperature, provided that this temperature is smaller than the
degradation temperature, $T_d$.  In the next section we shall devote especial
attention to this issue, since the literature does not seem to indicate any
previous studies of the influence of thermal effects in $y_c$ within the
models considered here.

\section{Critical Dopant  Density}

In this section we evaluate the critical dopant concentration, $y_c$, which,
with periodic boundary conditions in the polyacetylene chain, is given simply
as $y_c = a \rho_c$. In the next subsection we consider the case $T=0$
performing a numerical comparison between the MFA and the OPT using the
different sets of parameters presented earlier. Next, we will consider how the
temperature affects $y_c$.

\subsection{The zero temperature case}

In this section we will evaluate the dopant critical density in polyacetylene
within the OPT approach, neglecting eventual temperature effects which will be
considered in the next subsection.  Let us start by taking the limit $T \to 0$
in the free energy density, Eq. (\ref {Vdelta1}). The various functions
defined by Eqs.  (\ref{I1}), (\ref{I2}) and (\ref{I3}), in the $T \to 0$ limit
become

\begin{eqnarray}
&& \lim_{T\to 0} (kT)^2 I_1(\eta,\mu,T)=
- \frac{1}{2} \theta(\mu - \eta)
\left[ \eta^2 \ln \left( \frac{\mu + \sqrt{\mu^2 -
\eta^2}}{\eta}
\right) - \mu \sqrt{\mu^2 - \eta^2} \right]\;,
\label{I1T0}
\\
&&
\lim_{T \to 0} I_2(\eta,\mu,T) =  \theta(\mu-\eta)
\ln\left(\frac{\mu +\sqrt{\mu^2 - \eta^2}}{\eta} \right)\;,
\label{I2T0}
\\
&&
\lim_{T \to 0} kT I_3(\eta,\mu,T) =  {\rm sgn}(\mu) \theta(\mu-\eta)
\sqrt{\mu^2 - \eta^2} \;.
\label{I3T0}
\end{eqnarray}

\noindent
Using Eqs. (\ref{I1T0}), (\ref{I2T0}) and (\ref{I3T0}) in Landau's free energy
density, Eq. (\ref {Vdelta1}), we notice that it can be divided into two
cases: i) $\mu < \eta$ and ii) $\mu > \eta$. At zero temperature the critical
chemical potential, $\mu_c(0)$, is defined as the one which produces the same
pressure for both, ${\overline \Delta}_c = \Delta_0 \ne 0$ and ${\overline
  \Delta}_c=0$. This quantity, which has been evaluated in Ref. \cite{prdgn2},
is given by

\begin{equation}
\mu_c(0) = \frac{M}{\sqrt{2}} \exp
\left \{ -\frac {\pi}{\lambda_{\rm GN} \left ( 1- \frac{1}{2  N} \right )  }
\right \}
\left ( 1- \frac{\lambda_{\rm GN}}{2 \pi N} \right )^{-1/2}\;.
\label{muc}
\end{equation}
As expected, there are two values of $\rho$ corresponding to $\mu_c(0)$. The
first is simply $\rho=0$, corresponding to the minimum of the free energy
density that occurs at ${\overline \Delta}_c = \Delta_0 \ne 0$ for the case
$\mu < \eta$ (corresponding to the $T=\mu=0$ situation).  In the second case
($\mu > \eta$), the minimum of the free energy density occurs at the origin,
${\overline \Delta}_c=0$. In this case the PMS relation, Eq. (\ref{genpms}),
implies that $\bar \eta=0$ and one gets, from Eq. (\ref{density}) and after
simple algebra, the result

\begin{equation}
\rho_c(0)= \frac{\mu_c(0) N} {\pi \hbar v_F} 
\left ( 1- \frac{\lambda_{\rm GN}}{2 \pi N} \right ) \,,
\label{rhoc}
\end{equation}
where the multiplicative factor in the RHS of the above equation, $\mu_c(0)
N/(\pi \hbar v_F)$, is just the MFA result (with $N=2$). In Eq.  (\ref{rhoc})
the term $\lambda_{\rm GN}/(2 \pi N)$ gives the first order OPT finite $N$
corrections to $\rho_c(0)$.  One can now insert Eq.  (\ref{muc}) into Eq.
(\ref{rhoc}).  Using also the expression for $\Delta_0$, Eq. (\ref{deltaopt}),
we obtain an analytical expression for the critical density that includes
finite $N$ corrections,

\begin{equation}
y_c(0) = a \rho_c (0)  = \frac {\sqrt{2} N \Delta_0}{  \pi W}
\left ( 1- \frac{\lambda_{\rm GN}}{2 \pi N} \right )^{1/2}
\left ( 1- \frac{1}{2  N} \right ) \,\,,
\label{ycopt}
\end{equation}
where we have used the defining relation $a/(\hbar v_F) = 2/W$.  Note that the
explicit (overall) scale $M$ dependence disappears from Eq.  (\ref{ycopt}),
where only the physical parameters $W/\Delta_0$ set the overall scale.
However, there is an implicit scale dependence in the OPT case via the
dependence on $\lambda_{\rm GN}(M)$ within the factor $\left[ 1- \lambda_{\rm
    GN}/(2 \pi N) \right]^{1/2}$ \footnote{Note, however, that since $\left[1-
    \lambda_{\rm GN}/(2 \pi N) \right]^{1/2}[1-1/(2N)] \to 1$, as $N \to
  \infty$, the MFA results only depend on the ratio $2 \Delta_0 / W = E_{\rm
    gap}/W= a/\xi$.}.

Concerning our third prescription, the ``effective mean field" (EMF)
prescription III defined in Sec. IV, where the OPT corrections are
reinterpreted differently as redefining effective coupling and scale, the
corresponding expression of the critical dopant concentration is
straightforward to derive using definitions Eq.  (\ref{defeff}) and reads:
\begin{equation}
y^{EMF}_c(0) = \frac {\sqrt{2} N \Delta_0}{ \pi W^*_{EMF}}
\left[
1-\frac{\lambda^*_{EMF}}{2 \pi\: N \left ( 1- \frac{1}{2  N} \right ) } 
\right]^{1/2} \,\,,
\label{ycopteff}
\end{equation}
where it is again understood that corresponding set A or B data values should
be used now for $\lambda^*_{EMF}$ and $W^*_{EMF}$.
  
We are now in position to make predictions concerning the observable $y_c(0)$
using, for completeness, the two different sets of data parameter and our
three different theoretical prescriptions.  The comparison between the OPT and
MFA results is shown in Table II.

\begin{table}[pt]
{\begin{tabular}{c||c|c|c|c|c|c}
\hline 
 {}& I.A & I.B & II.A & II.B & III.A & III.B \\ \colrule \colrule
$y_c^{\rm OPT}$ & 0.0448 & 0.0576  &  0.0450 &  0.0569& 0.058 & 0.074\\ 
\colrule
$y_c^{\rm MFA}$  & 0.0630 & 0.081   &  0.0630 & 0.081& 0.0630 & 0.081\\ 
\colrule
$\xi/a$ & 7.143 & 5.555 & 7.143 & 5.555& 7.143 & 5.555 \\
\hline 
\end{tabular}}
\caption{The critical dopant concentration $y_c$  obtained with the OPT and the
MFA obtained from each of the three different prescriptions and two data sets 
at $T=0$.
{}For reference we also show the relevant values of
$\xi/a$ in each case.}
\end{table}

As one can see from Table II, the results depend rather substantially on the
experimental data set choice, due essentially to the (linear) dependence on
$\Delta_0$.  The most quoted experimental value of $y_c$ is $y_c\sim 0.06$,
although its precise value is not very accurately determined. Typically,
looking e.g. at the data from Refs. \cite{Fernando,moraes}, one may infer non
negligible uncertainties on the exact transition value, of about $\Delta y_c
\sim 0.01$. Also, slightly higher values of $y_c$ have been reported in other
studies \cite{conwell}.  In contrast, the two different prescriptions I and II
regarding the scale dependence only affect mildly the $y_c$ values in the OPT
case, {\it i.e.}, the arbitrary scale dependence, which appears only
indirectly within the factor $[ 1- \lambda_{\rm GN}/(2 \pi N) ]$, which
appears in our previous expressions like Eq. (\ref{ycopt}), remains moderate.

\begin{figure}[htb]
  \vspace{0.5cm}
  \epsfig{figure=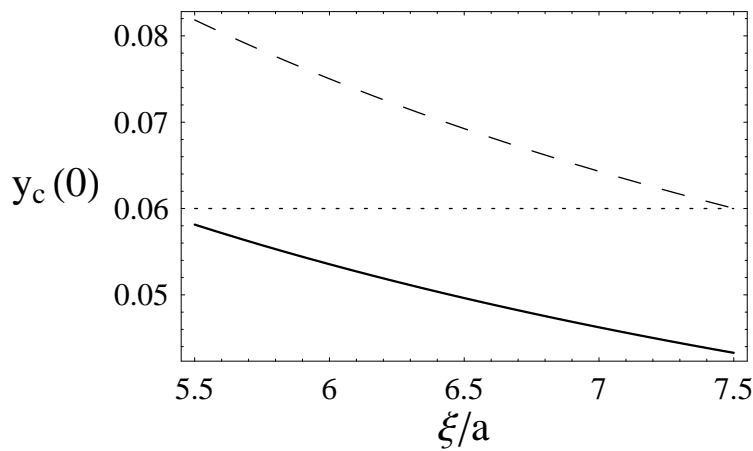,angle=0,width=10cm}
\caption[]{\label{allyc} The  OPT (continuous line) and MFA (dashed line) 
  predictions for $y_c(0)$ as a function of $\xi/a$, using prescription IB
  (Table \ref{tabdata}), in the relevant range $5.5 \leq \xi/a \leq 7.5$.  The
  experimental value is $y_c(0) \simeq 0.06$.}
\end{figure}

Inspection of Table II indicates that the OPT performs better for smaller
values of $\xi/a$. We have carried out numerical simulations that indicate
that, in fact, the OPT and MFA predict similar deviations from the
experimental value, $y_c =0.06$, for $\xi/a \simeq 6.4$.  {}For $\xi/a < 6.4$,
however, the OPT predictions are better than the MFA ones and the situations
gets reversed for $\xi/a > 6.4$ as Table II shows.  To illustrate this point
we present {}Fig. \ref {allyc}, where $y_c(0)$ is plotted as a function of
$\xi/a$, using prescription IB (Table \ref{tabdata}).  This figure shows that
for the value $\xi/a =6$, which Ref. \cite{campbell} refers to, the OPT ($y_c
= 0.0535$) performs better than the MFA ($y_c=0.0750$).  We also remark that
the same pattern is obtained when one considers others sets of values and
prescriptions.

{}Finally, as far the different prescriptions are concerned, we note that the
ones that make use of the band gap energy as $1.8$ eV (used in our data sets B
in Table I), produces results with much better agreement with the experimental
data for the OPT than the MFA, as one can check from the results in Table II.
In fact, this value for the band gap energy appears in the literature as a
more satisfactory value for polyacetylene \cite{Review}.

\subsection{The Finite Temperature Case}

\begin{figure}[htb]
  \vspace{0.5cm}
  \epsfig{figure=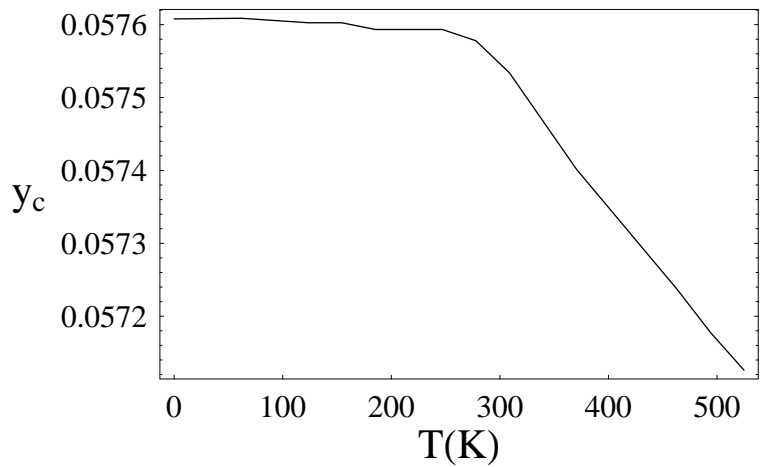,angle=0,width=10cm}
\caption[]{\label{yc} The critical dopant concentration, $y_c(T)$, temperature 
  dependence for parameters set B and procedure I. The degradation temperature
  is about $T_d \approx 400 K$.}
\end{figure}

To obtain the temperature dependence of the critical dopant concentration one
can proceed numerically considering $y_c(T)= a\rho_c (T)$. {}First, as already
emphasized, one should note that in practice there is an upper temperature of
about $T_d \sim 400 K$ above which our simple models break down since the
polymer undergoes other phase transitions.  {}Figure \ref{yc} shows that from
the absolute zero temperature to the upper temperature our prediction to the
decrease in the critical dopant concentration is only about $1\%$, while it is
less than $0.5 \%$ from room temperature (about $300 K$), where most of the
experiments are done, to the upper temperature. This shows that in practice,
at least with the type of models considered here, one may safely evaluate
$y_c$ at $T=0$ as it has been done in Refs. \cite{CM,CM2}.

\section{Conclusions}

Using the OPT we have reviewed the evaluation of Landau's free energy for the
1+1 dimensional massless Gross-Neveu model at finite temperature and chemical
potential as performed in Ref. \cite{prdgn2}. Then, relating this model to the
TLM continuous model for polyacetylene, we have computed the critical dopant
concentration, $y_c$, for the transition to the metallic phase.  The
(divergent) free energy density has been rendered finite by using a
renormalization procedure which is standard in quantum field theories
(${\overline {\rm MS}}$ scheme with dimensional regularization).  An arbitrary
energy scale, $M$, introduced during the formal regularization process was
fixed by using polyacetylene experimental inputs.  Regarding the matching of
the theory parameters to the experimental data values, we have provided a
critical discussion of the possible different prescriptions to relate them
giving, at the same time, the expected results for each of the prescriptions
used.

The OPT formalism allows for the inclusion of finite $N$ corrections already
at the first non trivial order which should improve the usual MFA results
since, for this particular case, $N=2$. To illustrate the possible phase
transitions allowed by the GN model we have obtained phase diagrams in the
$T-\mu$ as well as in the $T-\rho$ planes. Then, we have obtained a neat
analytical expression for $y_c= a\rho_c$ at $T=0$ which contains explicit
$1/N$ corrections. Our results show that when one uses up-to-date parameter
values, the OPT results improve over the MFA as expected and are in good
agreement with the experimental result. Another result of the present work
regards the study of possible thermal effects in $y_c$.  Our analysis has been
performed in a numerical fashion, showing that, for a realistic temperature
range, $0 < T < 400 \, K$, thermal effects induce a negligible decrease of
$y_c$ when going from the (mixed) semiconducting phase to the (symmetric)
metallic phase.  

Regarding the metal-insulator transition, one must recall
the importance of the transport property and the localization problem
\cite{localization,transport}.
As far as polymers are concerned, it is known that their transport properties
result from the mechanism of hopping \cite{transport}, which leads  the conductivity
to increase with the temperature till some maximum value. At the same time,
disorder in the system can result in the localization of states and,
if it is too strong, it can lead to an insulator behavior. The metal-insulator
transition in real systems is then a consequence of the interplay of 
the amount of disorder, doping and the thermal activation process.
On the theoretical side, an interesting possibility of extending the
field theory application method we used to study the metal-insulator
transition in polyacetylene, would be the calculation of the conductivity 
$\sigma$ and the determination of its $T,\mu$ dependence. This would then
allow a closer comparison with the experimental measures on this quantity.
Basically, from response theory, the conductivity can be computed from
a Green-Kubo formula, which entails a calculation of specific correlation
functions and higher loop Feynman diagram contributions in the GN 
model we used. Although the calculation of $\sigma$ in this context has 
been already considered within other approaches and approximations
in the literature, we hope to pursue a detailed 
calculation of $\sigma(T,\mu,N)$ including finite $N$ corrections, which we 
believe has not been done up to now. However, this is a non trivial
calculation which is well beyond the scope of the present paper. We 
intend to address this issue in the future.
Another interesting possible extension of the present work would
be to use the OPT and the methods developed in Ref. \cite{gnpolymers} to
consider the massive GN model, which can be related to the 
{\it cis}-polyacetylene.

\acknowledgments

H.C., M.B.P., and R.O.R. are partially supported by CNPq-Brazil.  H.C. also
thanks FAPEMIG for partial support.  We are grateful to Prof. A. Heeger for a
private communication regarding the experimental values of polyacetylene.  We
thank V. Mano for helpful information on thermal properties of polyacetylene
and M. Thies for sending us an important reference.


\begin{thebibliography}{99}
  
\bibitem{Shirakawa} H. Shirakawa, E. J. Louis, A. G. MacDiarmid, C. K. Chiang
  and A. J. Heeger, J. Chem. Soc., Chem. Commun., 578, 1977.
  
\bibitem{Nature1} J. H. Burroughes, C. A. Jones and R. H. Friend, Nature {\bf
    335}, 137, (1988).
  
\bibitem{Nature2} Y. Yu, M. Nakano and T. Ikeda, Nature {\bf 425}, 145 (2003).
  
\bibitem{Nature3} S. Kubatkin, A. Danilov, M. Hjort, J. Cornil, J.-L.  Bredas,
  N. Stuhr-Hansen, P. Hedega, and T. Bjørnholm, Nature {\bf 425}, 698 (2003).
  
\bibitem{Lin1} X. Lin, J. Li and S. Yip, Phys. Rev. Lett. {\bf 95}, 198303
  (2005).
  
\bibitem{Review} A. J. Heeger, S. Kivelson, J. R. Schrieffer and W. P. Su,
  Rev. Mod. Phys. {\bf 60}, 781 (1988).
  
\bibitem{Fernando} J. Chen, T. -C. Chung, F. Moraes and A. J. Heegerr, Solid
  State Commun. {\bf 53}, 757 (1985); F. Moraes, J. Chen, T. -C.  Chung and A.
  J. Heeger, Synth. Met. {\bf 11}, 271 (1985).
  
\bibitem{SSH} W. P. Su, J. R. Schrieffer and A. J. Heeger, Phys. Rev. Lett.
  {\bf 42}, 1698 (1979); Phys. Rev. B {\bf 22}, 2099 (1980).
  
\bibitem{Lin2} See, for example, X. Lin, J. Li, C. J. Forst and S. Yip, PNAS,
  {\bf 103}, 8943 (2006).
  
\bibitem{tlm} H. Takayama, Y.R. Lin-Liu and K. Maki, Phys. Rev B {\bf 21},
  2388 (1980).
  
\bibitem{Maxim} M. Mostovoy and J. Knoester, Phys. Rev. B {\bf 53}, 12057
  (1996).
\bibitem {CB} S. A. Brazoviskii and N. N. Kirove, JETP Lett. {\bf 33}, 4
  (1981); Pis'ma ZhETF {\bf 33}, 6 (1981); D. K. Campbell and A. R. Bishop,
  Phys. Rev.  {\bf B24}, 4859 (1981); Nucl. Phys. {\bf B200}, 297 (1982).
  
\bibitem{gn} D. Gross and A. Neveu, Phys. Rev. D {\bf 10}, 3235 (1974).
  
\bibitem{wolff} U. Wolff, Phys. Lett. {\bf B157}, 303 (1985).
  
\bibitem{fradkin}E. Fradkin and J. E. Hirsch, Phys. Rev. B {\bf 27}, 1680
  (1983).
  
\bibitem{otherrefs}A. Saxena and A. R. Bishop, Phys. Rev. A {\bf 44}, 2251
  (1991).
  
\bibitem{CM} A. Chodos and H. Minakata, Phys. Lett. {\bf A191}, 39 (1994).
  
\bibitem{CM2}A. Chodos and H. Minakata, Nucl. Phys. {\bf B490}, 687 (1997).
  
\bibitem{campbell} D. K. Campbell, Synt. Metals {\bf 125}, 117 (2002).
  
\bibitem{linear} A. Okopinska, Phys. Rev. D {\bf 35}, 1835 (1987); M. Moshe
  and A. Duncan, Phys. Lett. {\bf B215}, 352 (1988).
  
\bibitem{prdgn2} J.-L. Kneur, M. B. Pinto and R. O. Ramos, Phys.  Rev. D {\bf
    74}, 125020 (2006); Braz. J. Phys. {\bf 37}, 258 (2007).
  
\bibitem{Marino} E. C. Marino, Lizardo and H. C. M. Nunes, Nucl. Phys. {\bf
    B741}, 404 (2006), and references therein.
  
\bibitem {prdgn3}J.-L. Kneur, M. B. Pinto, R. O. Ramos and E. Staudt, Phys.
  Rev.  D {\bf 76}, 045020 (2007).
  
\bibitem{klimenko} K. G. Klimenko, Z. Phys. {\bf C37}, 457 (1988); B.
  Rosenstein, S. H. Park and B. J. Warr, Phys. Rev. D39, 3088 (1989); Phys.
  Rev. Lett.  {\bf 62}, 1433 (1989).
  
\bibitem {plb}J.-L. Kneur, M. B. Pinto, R. O. Ramos and E. Staudt, Phys. Lett.
  {\bf B657}, 136 (2007).
  
\bibitem {prb} F. F. Souza Cruz, M. B. Pinto and R. O. Ramos, Phys. Rev. B
  {\bf 64}, 014515 (2001); Laser Phys. {\bf 12}, 203 (2002).
  
\bibitem{pra} J.-L. Kneur, A. Neveu and M. B. Pinto, Phys. Rev. A {\bf 69},
  053624 (2004), J.-L. Kneur and M. B. Pinto, Phys. Rev. A {\bf 71}, 033613
  (2005); B. Kastening, Phys. Rev. A {\bf 70}, 043621 (2004).
  
\bibitem{prl}J.-L. Kneur, M. B. Pinto and R. O. Ramos, Phys. Rev. Lett.  {\bf
    89}, 210403 (2002); Phys. Rev. {\bf A68}, 043615 (2003).
  
\bibitem{braaten} E. Braaten and E. Radescu, Phys. Rev. Lett. {\bf 89}, 271602
  (2002); Phys. Rev. {\bf A66}, 063601 (2002).
  
\bibitem {landau} L. D. Landau and E. M. Lifshtiz, {\it Statistical Physics}
  (Pergamon, N.Y., 1958) p.  482.
  
\bibitem{mermin} N. D. Mermin and H. Wagner, Phys. Rev. Lett. {\bf 17}, 1133
  (1966).
  
\bibitem{coleman1} S. Coleman, Commun. Math. Phys. {\bf 31}, 259 (1973).
  
  
\bibitem{ma} R. F. Dashen, S.-K. Ma and R. Rajaraman, Phys. Rev. D {\bf 11},
  1499 (1974); S. H. Park, B. Rosenstein and B. Warr, Phys. Rept. {\bf 205},
  108 (1991).
  
\bibitem{gnpolymers} O. Schnetz, M. Thies and K. Urlichs, Ann. Phys. (NY) {\bf
    314}, 425 (2004); M. Thies and K. Urlichs, Phys. Rev. D {\bf 72}, 105008
  (2005); M. Thies, J. Phys. {\bf A39}, 12707 (2006).
  
\bibitem{pms} P. M. Stevenson, Phys. Rev. D {\bf 23}, 2961 (1981); Nucl. Phys.
  {\bf B203}, 472 (1982).
  
  
\bibitem{npb} S. K. Gandhi, H. F. Jones and M. B. Pinto, Nucl. Phys.  {\bf
    B359}, 429 (1991).

\bibitem{roth} T. Anderson and S. Roth, Braz. J. Phys. {\bf 24}, 746 (1994).
  
\bibitem{mixed}L. W. Shacklette and J. E. Toth, Phys. Rev. B {\bf 32}, 5892
  (1985).
  
\bibitem{moraes} T. C. Chung, F. Moraes, J. D. Flood and A. J. Heeger, Phys.
  Rev. B {\bf 29}, 2341 (1984).
  
\bibitem{conwell} E. M. Conwell, H. A. Mizes and S. Jeyadev, Phys. Rev. B {\bf
    40}, 1630 (1989).

\bibitem{localization}A. B. Kaiser, Rep. Prog. Phys. {\bf 64}, 1 (2001).

\bibitem{transport}V. N. Prigodin and A. J. Epstein, Synth. Met. {\bf 125}, 43 (2002).





\end{thebibliography}
\end{document}